
\magnification=1200
\overfullrule = 0 pt
\baselineskip=16 true bp

\def\ref#1{$^{#1}$}
\def\ea{{\it et al.}}
\def\ie{{\it i.e.}}

\hfuzz=0.5pt
\hbox{}
\centerline{\bf Diffusion-Limited Coalescence with Finite Reaction Rates
in One Dimension}
\vskip 1cm
\centerline{Dexin Zhong\footnote
*{bitnet: {\sl zhongd@craft.camp.clarkson.edu}}
and Daniel ben-Avraham\footnote
\ddag{bitnet: {\sl qd00@craft.camp.clarkson.edu}}}
\bigskip
\centerline{Clarkson Institute for Statistical Physics (CISP)}
\centerline{Physics Department, Clarkson University, Potsdam,
NY 13699-5820}

\vskip 1.2cm
\noindent{\bf Abstract:}

We study the diffusion-limited process $A+A\to A$ in one dimension, with finite
reaction rates.  We develop an approximation scheme based on the method of
Inter-Particle Distribution Functions (IPDF), which was formerly used for the
exact solution of the same process with infinite reaction rate.  The
approximation becomes exact in the very early time regime (or the
reaction-controlled limit) and in the long time (diffusion-controlled)
asymptotic limit.  For the intermediate time regime, we obtain a simple
interpolative behavior between these two limits.  We also study the coalescence
process (with finite reaction rates) with the back reaction $A\to A+A$, and in
the presence of particle input. In each of these cases the system reaches a
non-trivial steady state with a finite concentration of particles.  Theoretical
predictions for the concentration time dependence and for the IPDF are
compared to computer simulations.

\bigskip

\noindent PACS \#\ \  82.20.Mj,\ 02.50.+s,\ 05.40.+j,\ 05.70.Ln
\vfill\eject

\noindent{\bf 1. Introduction}
\medskip
In recent years much effort has been dedicated to Diffusion-limited
reactions in low dimensions.\ref{1-9}  Most research has focused on the
bimolecular reaction $A+B\to {\rm inert}$,\ref{2} and on
one-component coalescence, $A+A\to A$,\ref{3-7} and annihilation,
$A+A\to 0$.\ref{8,9} The last two systems were solved exactly in one
dimension and are especially useful in elucidating the anomalous kinetics
of diffusion-limited reaction processes.  Naturally, simple generalizations
and extensions of these basic processes are of much interest, because of the
possibility that they may also be solved exactly.  Indeed, the coalescence
process, $A+A\to A$, has been also solved together with the  back
reaction, $A\to A+A$, and under input of $A$ particles,\ref{5} as well as
for systems of finite size\ref{6} and with inhomogeneous initial
conditions.\ref{7}

In the diffusion-limited coalescence process reactions
occur at an infinite rate, taking place immediately upon the encounter of any
two particles. An obvious generalization would be to make the reaction rate
finite.  This more physical process would also exhibit an interesting crossover
from a classical, reaction-limited behavior, at early times, to a
diffusion-limited regime, in the long time asymptotic limit. Surprisingly,
this simple minded generalization makes it difficult (perhaps even
impossible) to solve the model exactly.

Simulations have been performed, both on the one-component annihilation process
and on the coalescence process in one dimension, with a finite reaction
rate.\ref{10-12} The simulations show three different regimes: ($a$) an early
time regime where the particles merely diffuse with a negligible change in the
initial concentration, ($b$) an intermediate regime, in which the concentration
of particles decays faster than the diffusion limited case, $c\sim 1/\sqrt{t}$,
but not quite as fast as the classical limit, $c\sim 1/t$, and ($c$) a long
time, diffusion-limited regime, where the system behaves exactly as if the
reaction rate were infinite. Based on these simulation results, the intriguing
possibility that in the intermediate regime the concentration decays
anomalously in time, with a  power dependent on the reaction rate, was
raised.\ref{10} Subsequent theoretical work by Privman \ea\ref{12} and
Hoyuelos and M\'artin\ref{11} has suggested that the kinetics in the
intermediate regime can be explained as an interpolation between the classical
and the diffusion-controlled limits.

The analysis of Privman \ea\ref{12} is based on the concept of inter-particle
distribution functions (IPDF), which had been introduced earlier for the exact
solution of the coalescence process with infinite reaction rate.\ref{5}
In the IPDF method a diffusion equation is derived for the probability that two
nearest particles are a distance $x$ apart (the IPDF), and the reaction is
represented by absorbing boundary conditions at the origin.  Privman \ea\ have
replaced this boundary condition with a radiative boundary condition, in order
to approximate the finite reaction rate.  Their approach yields an elegant,
{\it qualitative\/} understanding of the three regimes discussed above.
Hoyuelos and M\'artin\ref{11} have found an
interpolation formula for the challenging intermediate regime, which fits the
simulation data reasonably well.  However, the derivation is largely
phenomenological and it requires a scaling ansatz which we find physically
obscure.

In this paper, we study the one-dimensional coalescence process, $A + A \to A$,
with finite reaction rate.  Our approach is based on the IPDF method but
differs from that of Privman \ea\ in important details. In particular, it
allows
us to derive a closed analytic expression for the concentration decay at all
times.  For the intermediate regime, an interpolative formula between the
classical and diffusion-controlled limits emerges as a natural consequence of
the approximation. We also study the coalescence process
with back reactions,
$A\to A+A$,  and with particle input, $0\to A$.  In each of these two cases the
system arrives at a non-trivial steady state where the concentration is
finite.  Finally, we study the IPDF's themselves, which show different
characteristics at different time regimes and in the stationary cases.  Our
theoretical results compare  quite well with computer simulations.

The rest of the paper is organized as follows.  Our model and a physical
approximation based on the method of IPDF's are introduced in Section~2. In
Section~3, we perform a mathematical approximation that enables us to solve
the equations derived in Section~2 in closed form.  In the same section, we
compare between simulations and analytic results.  The two cases of
non-trivial steady state are explored in Section~4, both through our analytic
approach and through computer simulations.  We conclude with a discussion, in
Section~5.

\bigskip
\noindent{\bf 2. The reaction model and the IPDF method}
\medskip

Our model is defined on a one-dimensional lattice with lattice spacing
$\Delta x$.  Each site can
be either empty ($\circ$) or occupied ($\bullet$) with one particle.
The particles move randomly, independently of each other, to a nearest neighbor
site with a hopping rate $D/(\Delta x)^2$ (to each side).  On long length and
time scales this yields normal diffusion with diffusion coefficient $D$.  When
a particle hops onto a site which is occupied, coalescence takes place with
probability $k$, while with probability
$(1 - k)$ the particle is reflected back to its original position and no
reaction takes place. Thus, the probability $k$ controls the rate of the
reaction
$$
A + A \to A.                                                         \eqno(1)
$$
When $k=0$ the reaction rate is zero. The particles
merely diffuse, bouncing off each other and no reactions take place. In the
other extreme, when $k=1$ reactions are immediate, \ie, the reaction rate is
infinite.  This is the purely diffusion-limited case. When $k$ is very small
but finite, one expects to find a regime which is dominated by the slow
reaction
rate.  This is the classical, reaction-limited case.

To solve the system, we follow the IPDF method, previously used for the exact
solution of this model in the case of $k=1$.\ref{5}  We define, as usual,
$E_n(t)$, the probability that a randomly chosen segment of $n$ consecutive
sites is empty, \ie, contains no particles.  The probability that a site is
occupied is $1-E_1\equiv C$, and the density of occupied
sites is
$$
c(t)=(1-E_1(t))/\Delta x= C(t)/\Delta x.                               \eqno(2)
$$

$E_n$ gives the probability that, say, sites 1 through $n$ are empty,
while $E_{n+1}$ gives the probability that sites 1 through $n$ are empty and
also site $n+1$ is empty. Thus, the probability that a segment of $n$ sites
is empty, but that the adjacent $(n+1)$th site is occupied, is
${\rm Prob}(\circ\circ\cdots\circ\bullet)=E_n-E_{n+1}$. It can
also be shown\ref{5} that
$$
E_{n-1}-2E_n+E_{n+1}=Cp_n,                                         \eqno(3)
$$
where $p_n$ is the probability that nearest neighbor particles are exactly $n$
sites apart (this is the IPDF, from which the method derives its name). In
particular, the probability of finding two adjacent occupied sites is exactly
given by ${\rm Prob}(\bullet\bullet)=1-2E_1+E_2$.

When the reaction probability $k=1$, one can write down an exact equation for
the evolution of $E_n$:\ref{5}
$$
\partial_tE_n={2D\over(\Delta x)^2}\{(E_{n-1}-E_n)
               -(E_n-E_{n+1})\}.		                   \eqno(4)
$$
The first term on the r.h.s. describes the creation of an
empty $n$-sites interval, when a particle at the inner
edge of the interval hops out.  The second term describes annihilation of an
empty interval, when a particle at the outer edge of the interval hops in.

When the reaction probability $k<1$, hopping out of the interval is not always
possible. The target site may be occupied, in which case hopping (and
coalescence) will be disallowed, with probability $1-k$.  To account for this
effect, we require the probability of finding intervals of $n-1$ consecutive
empty sites, followed by two occupied sites. Since this probability cannot be
expressed in terms of the $E_n$ exactly, we propose the approximation
$$
{\rm Prob}(\overbrace{\circ\circ\cdots\circ}^{n-1}\bullet\bullet)\approx
{{\rm Prob}(\overbrace{\circ\circ\cdots\circ}^{n-1}\bullet)
{\rm Prob}(\bullet\bullet)\over{\rm Prob}(\bullet)}=
{(1-2E_1+E_2)(E_{n-1}-E_n)\over{1-E_1}}.                            \eqno(5)
$$
With this approximation, the evolution equation for arbitrary $k$ is
$$
\partial_tE_n={2D\over(\Delta x)^2}(E_{n-1}-2E_n+E_{n+1})
-{2D\over(\Delta x)^2}(1-k){{1-2E_1+E_2}\over{1-E_1}}{(E_{n-1}-E_n)},\eqno(6)
$$
where the correction due to failed coalescence attempts is represented by the
last term.  Notice that when $k=1$ this properly reduces to Eq.~(4).

Eq.~(6) is valid for $n>1$. For $n=1$ we have the exact equation:
$$
\partial_t E_1={2D\over(\Delta x)^2}k{(1-2E_1+E_2)},                 \eqno(7)
$$
which simply states that sites become empty at the same rate as
particles coalesce. Comparing Eqs.~(6) and (7) we see that they may be
combined, by requiring the boundary condition
$$
E_0(t)=1.                                                            \eqno(8)
$$
Eq.~(6) may thus be extended to the case of $n=1$.  A second boundary
condition is
$$
E_n(t)=0,\qquad {\rm as}\quad n\to\infty,                            \eqno(9)
$$
since an infinitely long segment will always contain particles, as long as the
concentration is finite.

The only approximation made here is in Eq.~(5), where some correlations
between the probable state of consecutive intervals are neglected.  We argue,
however, that Eq.~(6) is asymptotically correct both in the early and long
time regimes, and hence it may provide  a reasonable
interpolation for the intermediate time regime.  If the starting
configuration of the system is random, then $E_n=E_1^n$ and the state of
consecutive intervals is uncorrelated.  This situation will persist, and
Eq.~(5) will hold, until the concentration drop is noticeable (\ie, the
end of the early time regime) for it is only reactions that induce correlations
(in fact, diffusion randomizes the system). After very long times, on the other
hand, the concentration of particles becomes very small.  As a result, adjacent
occupied sites become extremely rare and the correction term in Eq.~(6)
eventually becomes negligible. Eq.~(6) then degenerates to the case of infinite
reaction rate and it does no longer matter how imprecise the correction term
is.

\bigskip
\noindent{\bf 3. Integration of the evolution equation}
\medskip

There exist various techniques to solve Eq.~(6), with the boundary conditions
of Eqs.~(8) and (9).  Perhaps the most straightforward method is numerical
integration, which requires discretization of the time variable in Eq.~(6). A
second approach, which worked well for the case of $k=1$, is passing to the
continuum limit.\ref{5} This is achieved by defining the spatial coordinate
$x=n\,\Delta x$. The probabilities $E_n(t)$ are replaced by the function
$E(x,t)$.  Letting $\Delta x\to 0$, Eq.~(6) is replaced by
$$
\partial_tE=2D\partial_x^2 E-2D(1-k){{\partial_x^2E|_{x=0}}
                 \over{\partial_xE|_{x=0}}}\partial_xE.            \eqno(10)
$$
and the boundary conditions of Eqs.~(8) and~(9) become
$$
E(0,t)=1,\qquad {\rm and} \qquad E(x\to\infty,t)=0.                \eqno(11)
$$
We see that the finite reaction rate gives rise to a non-local, non-linear term
in Eq.~(10). However, notice that
$\partial_x^2E|_{x=0}/\partial_xE|_{x=0}\equiv\omega(t)$ is a function of time
only.  Thus, in principle one can proceed by Laplace-transforming Eq.~(10)
with respect to the spatial variable $x$, and then determine $\omega(t)$
in some self-consistent way.  Unfortunately, this procedure leads to
complicated expressions and one is forced to resort to series expansions,
limiting the solution to a few asymptotic results.

Here we propose an alternative approach, based on an approximation of Eq.~(6).
We emphasize that this approximation is a mere {\it mathematical\/}
convenience,
designed to enable us to obtain a solution to Eq.~(6) in closed form.  The
{\it physical\/} approximation made in Eq.~(5) is the real focus of this
paper.  Indeed, Eq.~(6) can be solved to any degree of accuracy employing
numerical methods.

We first sum Eq.~(6) over the index $n$, from 1 to $\infty$, to yield
$$
\partial_t\sum_{n=1}^{\infty}E_n ={2D\over(\Delta x)^2}(1-E_1)
       -{2D\over(\Delta x)^2}(1-k){{1-2E_1+E_2}\over{1-E_1}},
                                                                    \eqno(12)
$$
where we have used the boundary conditions of Eqs.~(8) and (9). The r.h.s. can
be made a function of only $E_1$, with the help of Eq.~(7). For the l.h.s. we
make the approximation $\sum E_n\approx A/(1-E_1)$, where $A$ is a constant.
The motivation for this is that in the long time asymptotic limit
$E_n\approx 1$ for all $n$ up to a characteristic $\langle n\rangle=1/(1-E_1)$,
and falls sharply to zero for $n>\langle n\rangle$.  More precisely, in the
long
time asymptotic limit the reaction proceeds as if $k$ is effectively 1, in
which case we know that
$A=2/\pi$, exactly.\ref{5} The approximation lies in the fact that we assume
$A$ to be constant {\it at all times}.  Indeed, the variation in
$A$ is quite small; at the beginning of the process, when the distribution is
random, $A=1$. Let us then assume $A=2/\pi$ (to match the exact long time
asymptotic solution) to hold true throughout the process.  Eq.~(12) then
becomes
$$
{d\over d\tau}{2\over\pi C}=C+{1-k\over kC}{d C\over d\tau},
                                                                    \eqno(13)
$$
where $\tau\equiv (2D/(\Delta x)^2)t$ is a dimensionless time variable, and
we have used $1-E_1=C$ (Eq.~2).
The solution to Eq.~(13) is
$$
C={1-k+\sqrt{({2k\over\pi C_0}+1-k)^2+{4k^2\over\pi}\tau}  \over
    2({k\over\pi C_0^2}+{1-k\over C_0}+k\tau)},                     \eqno(14)
$$
where $C_0\equiv C(t=0)$.

In Fig.~1, we plot the
concentration decay as computed from a numerical integration of Eq.~(6), from
the analytical expression of Eq.~(14), and from computer simulations, for the
same choice of parameters ($D$ and $C_0$) and for various choices of $k$. The
agreement between Eq.~(14) and the numerical integration is excellent---little
is lost in the `mathematical' approximation.  More importantly, there is good
agreement between theory and simulations:  the early and late time regimes
match almost perfectly, and in spite of differences of up to 9~\% in the
intermediate regime, the slope of the curves and the crossover times are almost
identical.

{}From Eq.~(14), we see that the intermediate time regime is merely an
interpolation between a classical decay, $C\sim 1/t$, and a diffusion-limited
decay, $C\sim 1/\sqrt{t}$. We can use our result to estimate the crossover
times.  We first expand $C(\tau)$ in powers of $t$:
$$
C(\tau)=C_0-{\pi kC_0^3\over 2k+\pi(1-k)C_0}\tau+{\cal O}(\tau^2).  \eqno(15)
$$
We obtain the crossover time between the early time regime and the
intermediate regime, $\tau_1$, by requiring that the term linear in $\tau$ be a
finite fraction, $\epsilon$, of $C_0$:
$$
\tau_1=\epsilon({2\over\pi C_0^2}+{1-k\over kC_0}).                 \eqno(16)
$$
Even in the diffusion-limited case, when $k\approx 1$, there is an early time
regime where reactions go unnoticed.  Since coalescence is immediate, the
crossover time equals the typical time that two nearest particles will take to
reach each other. The average distance between particles is $1/c_0$, and since
the particles diffuse, $t_1\approx 1/Dc_0^2$,\ref{13} in agreement with
the estimate above.  The second term on the r.h.s. predicts that the crossover
time $\tau_1$ will increase proportionally to $1/kc_0$, which is
characteristic of the classical limit (classically, $dc/dt=-kc^2$ and
$c=c_0/(1+kc_0t)$).

Next, we expand Eq.~(14) in powers of $1/\sqrt{\tau}$,
$$
C(\tau)={1\over\sqrt{\pi\tau}}+{1-k\over 2k\tau}+{\cal O}({1\over \tau^{3/2}}).
                                                                   \eqno(17)
$$
Notice that the leading term corresponds to the long time
asymptotic limit, where the concentration decays as $c=1/\sqrt{2\pi Dt}$ (as in
the exact solution for $k=1$).  Interestingly, the second term, too, does not
retain any memory of the initial density, $C_0$, but has some $k$-dependence.
Comparing the leading term to the first correction we get an estimate for
$\tau_2$, the crossover time between the intermediate regime and the long time
regime:
$$
\tau_2={\pi(1-k)^2\over 4k^2}.                                     \eqno(18)
$$
This can be explained heuristically, as follows. The long time asymptotic
regime occurs because the reaction probability $k$ effectively renormalizes
to~1: If the density of particles is $C$, the number of sites between
neighboring particles is on the average $1/C$.  Because the particles
diffuse, it takes them of the order of $1/C^2$ steps to meet each other. During
this time, each of the $1/C$ sites is visited of the order of
$(1/C^2)/(1/C)=1/C$ times.  In particular, two neighboring particles will
collide about $1/C$ times before wandering off each other to interact with
other particles. This means that two neighboring particles will react almost
surely, before meeting other partners, if $(1/C)k\approx 1$. That is, the
crossover will occur when
$C\approx k$, or, since in the long time asymptotic
regime $C=1/\sqrt{\pi\tau}$, $\tau_2\approx 1/\pi k^2$, in agreement
with Eq.~(18).

To obtain the IPDF, we integrate Eq.~(6) numerically and then use Eq.~(3).  At
time $t=0$, we start with a random distribution of particles, so that
$p_n(0)=(1-C_0)^n/C_0$.  In the continuum approximation, this can be written
in the scaling form $p(\xi)=\exp(-\xi)$, where $\xi\equiv c(t)x$. As the
reaction proceeds, the likelihood to find nearby particles decreases, due to
the coalescence process.  Thus, coalescence gives rise to an effective
repulsion. In the long time asymptotic limit, when
$k$ effectively renormalizes to 1, the IPDF arrives at the stationary scaling
form $p(\xi)=(\pi\xi/2)\exp(-\pi\xi^2/4)$.\ref{5} During the intermediate time
regime, the IPDF makes a smooth transition between these two limits.  In
Fig.~(2), we plot $p(\xi)$ as obtained from numerical integration and compared
to computer simulations at various stages of the process. As might
be expected, the agreement is worst in the intermediate time regime.

\bigskip
\noindent{\bf 4. Coalescence with non-trivial steady states}
\medskip

Until now, we have treated the case where in the long time asymptotic limit
the concentration drops to zero.  We now want to discuss situations with a
non-trivial steady state.  It is important to examine such situations because
they impose a stricter test on the validity of our approximation.

\medskip
\noindent{\sl Back reactions}
\medskip

We consider first the case where the reaction (1) is reversible:\ref{5}
$$
    A+A \rightleftharpoons A.                                      \eqno(19)
$$
In the reverse process, $A\to A+A$, a particle gives birth to another at an
adjacent site, at rate $v/\Delta x$ (this means rate $v/2\Delta x$ on either
side of the original particle).  The corresponding evolution equation is
$$
\eqalign{\partial_t E_n=
&{2D\over(\Delta x)^2}(E_{n-1}-2E_n-E_{n+1})
               -{2D\over(\Delta x)^2}(1-k)
               {{1-2E_1+E_2}\over{1-E_1}}{(E_{n-1}-E_n)} \cr
            &  -{v\over {\Delta x}}{(E_n-E_{n+1})},\cr}            \eqno(20)
$$
where the last term describes the annihilation of an $n$-sites empty
interval due to a birth event from a particle at its outer edge.

In the steady state, the l.h.s. is equal to zero and Eq.~(20) becomes a
recursion relation for the $E_n$.  The solution can be
found by assuming a steady state of maximum entropy: since the process is
reversible, the steady state is an {\it equilibrium\/} state and hence
the IPDF is a Poisson distribution, $E_n=E_1^n$. Taking into account the
boundary conditions of Eqs.~(8) and (9), we find
$$
E_n=\big({2Dk\over  2Dk+v\Delta x}\big)^n.                         \eqno(21)
$$
and
$$
c_s={1-E_1\over\Delta x}={v\over 2Dk+v\Delta x},                     \eqno(22)
$$
Notice that although Eq.~(20) contains an approximation, the corresponding
steady-state equation is exact:  when the IPDF is completely random the
approximation of Eq.~(5) becomes exact.  Thus, the result of Eqs.~(21) and
(22) is exact.  This is well confirmed by simulations.

\medskip
\noindent{\sl Particles input}
\medskip

Consider now the case of a random, steady input of particles. At each time
step empty sites become occupied with probability $R\Delta x$.  That is,
$R$ is the increase in concentration per unit time, due to input.\ref{5} The
evolution equation is
$$
\eqalign{\partial_t E_n=
&{2D\over(\Delta x)^2}(E_{n-1}-2E_n-E_{n+1})
               -{2D\over(\Delta x)^2}(1-k)
               {{1-2E_1+E_2}\over{1-E_1}}{(E_{n-1}-E_n)} \cr
            &  -Rn\Delta x E_n,\cr}                               \eqno(23)
$$
where the last term represents the annihilation of an $n$-sites empty
interval due to input.

Here the steady state limit is less simple than for back reactions.  Although
the input sustains a steady concentration of particles at the long
time asymptotic limit, this stationary state is not a true equilibrium state
(the input process is {\it not\/} the reverse of coalescence) and the
particles are more ordered than in a Poisson's distribution.  Nevertheless,
Eq.~(23) with $\partial_tE_n=0$ still yields an approximate recursion relation
for the stationary $E_n$.  This can be solved exactly, but the solution is
rather cumbersome.  Instead, it is more enlightening to consider the continuum
limit of the steady state equation:
$$
0=2D\partial_x^2 E-2D(1-k)\omega\partial_xE-xRE,                  \eqno(24)
$$
where
$\omega=\partial_x^2E|_{x=0}/\partial_xE|_{x=0}=-\partial_x^2E|_{x=0}/c_s$ is
now a constant. To determine $\omega$, let us look at the discrete steady state
equation for $n=1$,
$$
0={2D\over(\Delta x)^2}k(1-2E_1+E_2)-\Delta x  R E_1.             \eqno(25)
$$
It simply equates the rates of input events and coalescence
events in the steady state.  The continuum limit of Eq.~(25) is too drastic in
that it yields zero for each of these rates (and also $\omega=0$).

A somewhat inelegant,
but effective way around this is to retain $\Delta x R$ finite, so that
$\omega=-R\Delta x/2Dkc_s\neq 0$.  Then, the solution to Eq.~(25) with the
boundary conditions of Eq.~(11) is
$$
E(x)=\exp(-{x\over\kappa c_s})\,{{\rm Ai}[r^{1/3}x+r^{-2/3}/(\kappa c_s)^{2}]
                       \over{\rm Ai}[r^{-2/3}/(\kappa c_s)^{2}]},    \eqno(26)
$$
where $\kappa=4Dk/(1-k)R\Delta x$ and $r=R/2D$.  From the relation
$c_s=-\partial_xE|_{x=0}$, we then obtain a transcendental equation for the
steady state concentration, $c_s$:
$$
c_s={1\over\kappa c_s}-r^{1/3}{{\rm Ai}'[r^{-2/3}/(\kappa c_s)^{2}]
                       \over{\rm Ai}[r^{-2/3}/(\kappa c_s)^{2}]}.    \eqno(27)
$$

In Fig.~3, we plot $c_s$ as a function of the reaction probability $k$,
for fixed $r$ ($r^{1/3}=0.04$), as obtained from computer simulations and from
numerical integration of the discrete steady state equation. The agreement
between simulations and theory is quite good. Also, for the range shown, the
agreement between Eq.~(27) and the numerical integration is better than 4~\%.

{}For small concentrations,
Eq.~(27) itself may be simplified. If $c\ll r^{1/3}/\kappa$, then
Ai$'[r^{-2/3}/(\kappa c_s)^2]/$Ai$[r^{-2/3}/(\kappa c_s)^2]\approx
{\rm Ai}'(0)/{\rm Ai}(0)= 0.72901\dots$, and the equation reduces to a simple
quadratic, with the solution
$$
c_s={1\over 2}\Big( \tilde c_s
    +\sqrt{\tilde c_s^2+{2(1-k)\Delta x\over k} r}\,\,\,\Big),\qquad
\tilde c_s=-{{\rm Ai}'(0)\over{\rm Ai}(0)}r^{1/3}.                  \eqno(28)
$$
Here $\tilde c_s$ is the steady state concentration when $k=1$, and is
exact.\ref{5} In the range shown in Fig.~3, Eq.~(28) agrees with the $c_s$
obtained from numerical integration of Eq.~(23) to within 10~\%.
We emphasize, again, that the mathematical approximations made here are not
essential to our method, but are merely done to obtain  simple final
expressions, to better understand the consequences of the physical
approximation made in writing Eq.~(23).

Finally, the IPDF itself is obtained from
$p(x)=(1/c_s)\partial^2E/\partial x^2$, and using Eq.~(26) with the $c_s$ found
from numerical integration (or any of the approximation formulae).  In Fig.~4,
we compare between the IPDF's obtained in this way and from computer
simulations for $r^{1/3}=0.04$ fixed, and various values of $k$.  The tail of
$p(x)$ falls off as $\exp(-x^{3/2})$ and for moderately large $k$ $p(x)$ shows
a maximum. Thus, the system is more ordered than in the case of back
reactions, but less ordered than in the case of pure coalescence, where the
tail of the IPDF decays as $\exp(-x^2)$. The agreement between theory and
simulations is best for large $k$. From Figs.~3 and 4 one can see that the
IPDF is a much more sensitive test for approximations than the concentration
alone.

An interesting result is that
$p(x=0)$ is zero only when $k=1$. This is in contrast to
the case of pure coalescence, where in the long time asymptotic limit
$p(0)=0$, regardless of the value of $k$.  Recall that in the latter case $k$
effectively renormalizes to 1 when the concentration decreases.  This
cannot happen in the case of back reactions, when $c$ approaches a stationary
{\it finite\/} value.  Then,  if the reaction rate is small enough it can
overcome the strong effective repulsion between nearest particles.

\bigskip
\noindent{\bf 5. Summary and discussion}
\medskip

We have introduced an approximation, based on the IPDF method, that allows us
to draw analytically simple descriptions of the coalescence process $A+A\to A$
in one dimension, with finite reaction rates.  The intermediate time regime is
convincingly shown to be a crossover behavior between the classical,
reaction-controlled limit, in which the concentration decays as $c\sim 1/t$,
and the diffusion-controlled limit, where $c\sim 1/\sqrt{t}$.  This result,
summarized in Eq.~(14), is similar to that of Hoyuelos and M\'artin.\ref{11} It
requires no fitting parameters and no approximations other than the standard
approximation of Eq.~(5).

The approximation yields exact results for the equilibrium state, reached in
the presence of the reverse reaction $A\to A+A$.  This is because in the
equilibrium state the fundamental approximation of Eq.~(5) becomes exact.
Surprisingly, we obtain rather accurate results for the stationary state of the
process with input, in spite of the strong spatial correlations that do
evolve.

The present approach is
similar to that of Privman \ea\ref{12} Their equation for the IPDF,
$$
\partial_t p(x,t)=2D\partial_x^2p(x,t)+2D(1-k)p(0,t)\partial p(x,t), \eqno(29)
$$
is closely related to our Eq.~(10), through $p(x,t)=\partial_x^2E(x,t)/c(t)$.
The difference is that ${\rm
Prob}(\circ\circ\cdots\circ\bullet\bullet)\approx{\rm
Prob}(\circ\circ\cdots\circ\bullet){\rm Prob}(\bullet)$ is used, instead of
our Eq.~(5).  Although Eq.~(5) is a bit more sensitive to short range
correlations, we believe that this difference is trivial.  The
advantage of our approach is more likely in that in writing an evolution
equation for
$E(x,t)$, we are able to use the exact boundary condition $E(0,t)=1$. In
contrast, there is no such constraint on $p(x=0,t)$ of Eq.~(29).

Regarding reverse reactions and input of particles, we have discussed here only
the stationary limits and have not addressed the time-dependent relaxation to
the steady-state.  A straightforward separation of variables and decomposition
into eigenvalue equations is not possible, because of the non-linear nature of
the correction term arising from the finite reaction-rate. In the case of back
reactions and infinite reaction-rate, there is a dynamical phase transition in
the characteristic relaxation time.\ref{5}  It will be interesting to see how
is this transition affected by finite reaction rates.  Is there a range of
reaction-rates for which the transition disappears?  The approximation method
used here has been successfully employed, with slight modifications, for the
study of the diffusion-limited processes $3A\to 2A$ and $3A\to A$,\ref{14} and
for the contact process in one dimension.\ref{15} It may be worthwhile to
further explore its potential. For example, Eq.~(5) can be systematically
improved, by writing an hierarchy of evolution equations for the state
probabilities of finite size intervals and introducing a truncation scheme at a
later stage.  These intriguing questions are left for future work.

\medskip
\noindent{\bf Acknowledgments}
\smallskip
We thank Ely Ben-Naim for his help in the preliminary stages of this research.
We also thank Ely Ben-Naim, Charlie Doering, Paul Krapivsky, and Vladimir
Privman for numerous, useful discussions.

\vfill\eject
\centerline{\bf References}
\medskip

\item{[1]} See, for example, {\sl J. Stat. Phys.} {\bf 65}, nos.~5/6 (1991):
this issue contains papers presented at a conference on Models of Non-Classical
Reaction Rates, which was held at NIH (March 25--27, 1991) in honor of
the 60th birthday of G.~H.~Weiss.
\item{[2]} M. Bramson and J. L. Lebowitz, {Phys. Rev. Lett.} {\bf 61},
2397 (1988) and references therein.
\smallskip
\item{[3]} M. Bramson and D. Griffeath, {\sl Ann. Prob.} {\bf 8}, 183 (1980).
\smallskip
\item{[4]} D. C. Torney and H. M. McConnell, {\sl J. Phys. Chem.}
{\bf 87}, 1941 (1983).
\smallskip
\item{[5]} C.~R.~Doering and D.~ben-Avraham, {\sl Phys. Rev. Lett.}
{\bf 62}, 2563 (1989); M.~A.~Burschka, C.~R.~Doering, and
D. ben-Avraham, {\sl Phys. Rev. Lett.} {\bf 63}, 700 (1989);
D. ben-Avraham, M. A. Burschka, and C. R. Doering, {\sl J. Stat. Phys.}
{\bf 60}, 695 (1990).
\smallskip
\item{[6]} C.~R.~Doering and M.~A.~Burschka, {\sl Phys. Rev. Lett.} {\bf 64},
245 (1990).
\smallskip
\item{[7]} C.~R.~Doering, M.~A.~Burschka, and W.~Horsthemke, {\sl J. Stat.
Phys.} {\bf 65}, 953 (1991)
\smallskip
\item{[8]} Z. Racz, {\sl Phys. Rev. Lett.} {\bf 55}, 1707 (1985).
\smallskip
\item{[9]} A. A. Lushnikov, {\sl Phys. Lett.} {\bf 120A}, 135 (1987).
\smallskip
\item{[10]} L. Braunstein, H.~O.~M\'artin, M.~D.~Grynberg, and H.~E.~Roman,
{\sl J. Phys. A} {\bf 25}, L255 (1992).
\smallskip
\item{[11]} M.~Hoyuelos and H.~O.~M\'artin, {\sl Rate equation of the
$A+A\to A$ reaction with probability of reaction and diffusion}, preprint.
\smallskip
\item{[12]} V.~Privman, C.~R.~Doering, and H.~L.~Frisch, {\sl Phys. Rev. E}
{\bf 48}, 846 (1993).
\smallskip
\item{[13]} K.~Kang and S.~Redner, {\sl Phys. Rev. A} {\bf 32}, 435 (1985).
\smallskip
\item{[14]} D. ben-Avraham, {\sl Phys. Rev. Lett.} {\bf 71}, 3733 (1993);
D.~ben-Avraham and D.~Zhong, {\sl Chem. Phys.} {\bf 180}, 329 (1993).
\smallskip
\item{[15]} E.~Ben-Naim and P.~L.~Krapivsky, {\sl Cluster approximation
for the contact process}, preprint.

\vfill\eject
\centerline{\bf Figure Captions}
\bigskip
{\parindent 0cm

{\bf Figure 1:} Concentration decay as a function of time for the coalescence
process as obtained from computer simulations (solid line), and from Eq.~(14)
(broken line). The different curves represent different values of the reaction
probability; $k=0.005$, 0.01, 0.02, 0.04, 0.08, and 0.16 (from top to bottom).
Shown also are results from numerical integration of Eq.~(6) for $k=0.02$
(circles).
\medskip

{\bf Figure 2:} The scaled IPDF, $p(cx)$, for the coalescence process for
$k=0.02$, as obtained from computer simulations (circles) and from numerical
integration of Eq.~(6) (solid line).  Results are shown for $t=0$, 100, 1000,
and 10000 ($p(0)$ decreases with time) .  The crossover times for this
particular process are roughly at
$t_1=15$ and $t_2=4000$. Shown also is the IPDF in the long time asymptotic
limit (broken line).
\medskip

{\bf Figure 3:} The stationary concentration, $c_s$, as a function of the
reaction probability, $k$ for coalescence with input of particles, with
$r=(0.04)^3$.  Numerical integration of Eq.~(23) (circles) is compared to
computer simulation results (solid line).
\medskip

{\bf Figure 4:} The scaled IPDF, $p(c_sx)$ for the same process as in Fig.~3
for
$k=0.02$, 0.04, 0.08, 0.20, and 0.40 ($p(0)$ is smaller for larger $k$), as
computed from the second derivative of Eq.~(26) (broken line) and compared to
computer simulations (solid line).

}\bye